# Precise and ultrafast molecular sieving through graphene oxide membranes

R. K. Joshi[1], P. Carbone[2], F. C. Wang[3], V. G. Kravets[1], Y. Su[1], I. V. Grigorieva[1], H. A. Wu[3], A. K. Geim[1*], R. R. Nair[1*]

[1]School of Physics & Astronomy, University of Manchester, Manchester, M13 9PL, UK
[2]School of Chemical Engineering & Analytical Science, Manchester, M13 9PL, UK
[3]CAS Key Laboratory of Mechanical Behavior and Design of Materials, Department of Modern Mechanics, University of Science and Technology of China, Hefei, Anhui 230027, China

*There has been intense interest in filtration and separation properties of graphene-based materials that can have well-defined nanometer pores and exhibit low frictional water flow inside them. Here we investigate molecular permeation through graphene oxide laminates. They are vacuum-tight in the dry state but, if immersed in water, act as molecular sieves blocking all solutes with hydrated radii larger than 4.5Å. Smaller ions permeate through the membranes with little impedance, many orders of magnitude faster than the diffusion mechanism can account for. We explain this behavior by a network of nanocapillaries that open up in the hydrated state and accept only species that fit in. The ultrafast separation of small salts is attributed to an 'ion sponge' effect that results in highly concentrated salt solutions inside graphene capillaries.*

Porous materials with a narrow distribution of pore sizes, especially in an angstrom range (*1-5*), attracts special attention because of possible applications in filtration and separation technologies (*5-7*). The observation of fast permeation of water through carbon nanotubes (*8-10*) and, more recently, through graphene-oxide (GO) laminates (*11*) has resulted in many proposals to use these materials for nanofiltration and desalination (*8-19*). GO laminates are particularly attractive because they are easy to fabricate, mechanically robust and offer no principal obstacles towards industrial scale production (*20,21*). They are made of impermeable functionalized graphene sheets that have a typical size $L \approx 1$ μm and the interlayer separation, $d$, sufficient to accommodate a mobile layer of water (*11-25*).

We have studied GO laminates that were prepared from GO suspensions by using vacuum filtration as described in Supporting Material (*25*). The resulting membranes were checked for their continuity by using a helium leak detector before and after filtration experiments, which proved that the membranes were vacuum-tight in the dry state (*11*). Figure 1 shows schematics of our experiments. First, we have filled the feed container with various liquids including water, glycerol, toluene, ethanol, benzene and dimethyl sulfoxide (DMSO) and monitored leaks into the permeate container that is left either empty or contains a different liquid. No permeation could be detected over a period of many weeks by monitoring liquids' levels and using chemical analysis (see below). The behavior agrees with the previously reported properties of GO laminates that block all liquids and gases except for water (*11*). In addition, this shows that the presence of water in only one of the containers and a miscible liquid in the other (for example, glycerol) is insufficient to open up capillaries through the entire membrane thickness and allow interdiffusion.

The situation changes if both containers are filled with water solutions. In this case, permeation through the same vacuum-tight membrane can readily be observed as rapid changes in liquid levels (several mm per day for the setup in Fig. 1B). The direction of flow is given by osmosis. For example, a level of a sucrose solution in the feed container rises whereas it falls in the permeate container filled with deionized water. For a membrane with a thickness $h$ of 1 μm, we find water flow rates of ≈0.2 L m$^{-2}$ h$^{-1}$ for 1 M feed solutions, and the speed increases with increasing the molar concentration $C$. Because 1 M corresponds to an osmotic pressure of ≈25 bar at room temperature, the flow rates agree with the evaporation rates of ≈10 L m$^{-2}$ h$^{-1}$ reported for similar GO membranes, in which case the permeation was driven by a capillary pressure of the order of 1,000 bars (*11*).

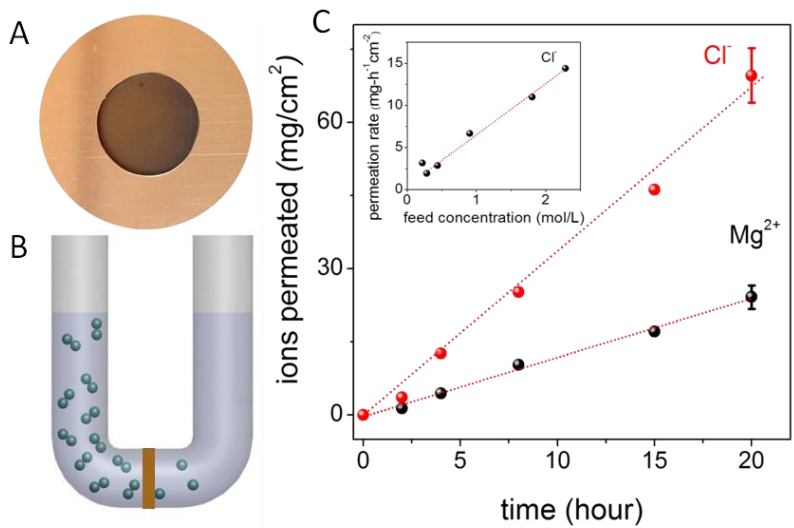

*Fig. 1. Ion permeation through GO laminates. (**A**) Photograph of a GO membrane covering a 1 cm opening in a copper foil. (**B**) Schematic of the experimental setup. The membrane separates the feed and permeate containers (left and right, respectively). Magnetic stirring is used to ensure no concentration gradients. (**C**) Filtration through a 5 μm thick GO membrane from the feed container with a 0.2 M solution of MgCl$_2$. The inset shows permeation rates as a function of C in the feed solution. Within our experimental accuracy (variations by a factor of <40% for membranes prepared from different GO suspensions), chloride rates were found the same for MgCl$_2$, KCl and CuCl$_2$. Dotted lines are linear fits.*

After establishing that GO membranes connect the feed and permeate containers with respect to transport of water molecules, we have investigated the possibility that dissolved ions and molecules can simultaneously diffuse through capillaries. To this end, we have filled the feed container with various solutions and studied if any of the solutes appears on the other side of GO membranes, that is, in the permeate container filled with deionized water (Fig. 1B). As a quick test, ion transport can be probed by monitoring electrical conductivity of water in the permeate container (*Fig. S1*). We have found that for some salts (for example, KCl) the conductivity increases with time but remains unaffected for others (for example, K$_3$[Fe(CN)$_6$]) over many days of measurements. This suggests that only certain ions may diffuse through GO laminates. Note that ions are not dragged by the osmosis-driven water flow but move in the opposite direction.

To quantify permeation rates for diffusing solutes and test those that do not lead to an increase in conductivity (sucrose, glycerol and so on), we have employed various analytical techniques. Depending on a solute, we have used ion chromatography, inductively coupled plasma optical emission spectrometry, total organic carbon analysis and optical absorption spectroscopy (*25*). As an example, Figure 1C shows our results for MgCl$_2$ which were obtained by using ion chromatography and inductively coupled plasma optical emission spectrometry for Mg$^{2+}$ and Cl$^-$, respectively. One can see that concentrations of Mg$^{2+}$ and Cl$^-$ in the permeate container increase linearly with time, as expected. Slopes of such curves yield permeation rates. The inset of Fig. 1C illustrates that the observed rates depend linearly on $C$ in the feed container. Note that cations and anions move through membranes in stoichiometric amounts so that charge neutrality within each of the containers is preserved. Otherwise, an electric field would build up across the membrane, slowing fast ions until the neutrality is reached. In Fig. 1C, permeation of one $Mg^{2+}$ ion is accompanied by two ions of chloride, and the neutrality condition is satisfied.

Figure 2 summarizes our results obtained for different ionic and molecular solutions. The small species permeate with approximately the same speed whereas large ions and organic molecules exhibit no detectable permeation. The effective volume occupied by an ion in water is characterized by its hydrated radius. If plotted as a function of this parameter, our data are well described by a single-valued function with a sharp cutoff at ≈4.5Å (Fig. 2). Species larger than this are sieved out. This behavior corresponds to a physical size of the mesh of ≈9Å. Fig. 2 also shows that permeation rates do not exhibit any notable dependence on ion charge (*12,13,23,26*) and triply charged ions such as $AsO_4^{3-}$ permeate with approximately the same rate as singly-charged $Na^+$ or $Cl^-$. Finally, to prove the essential role of water for ion permeation through GO laminates, we dissolved KCl and $CuSO_4$ in DMSO, the polar nature of which allows solubility of these salts. No permeation has been detected, proving that the special affinity of GO laminates to water is important.

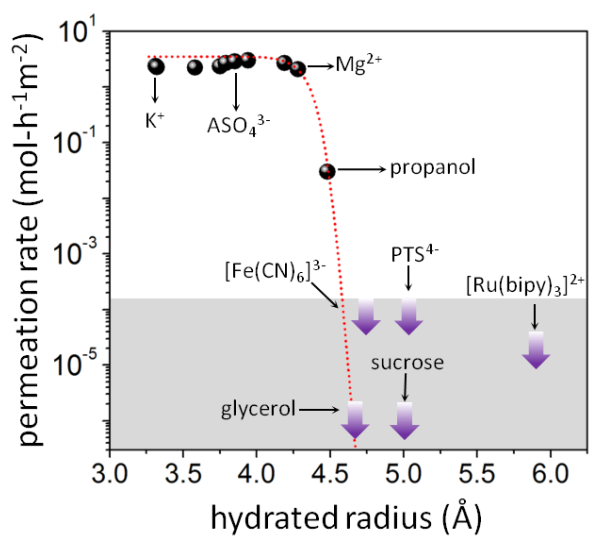

*Fig. 2. Sieving through an atomic scale mesh. The shown permeation rates are normalized per 1M feed solution and measured by using 5 μm thick membranes. Some of the tested chemicals are named here; the others can be found in Table S1 (25). No permeation could be detected for the solutes shown within the grey area during measurements lasting for 10 days or longer. The thick arrows indicate our detection limit that depends on a solute. Several other large molecules including benzoic acid, DMSO and toluene were also tested and exhibited no detectable permeation. The dashed curve is a guide to the eye, showing an exponentially sharp cutoff with a semi-width of ≈0.1Å.*

To explain the observed sieving properties, we employ the model previously suggested to account for unimpeded evaporation of water through GO membranes (*11*). Individual GO crystallites have two types of regions: functionalized (oxidized) and pristine (*21,27,28*). The former regions act as spacers that keep adjacent crystallites apart. In a hydrated state, the spacers help water to intercalate between GO sheets, whereas the pristine regions provide a network of capillaries that allow nearly frictionless flow of a layer of correlated water, similar to the case of water transport through carbon nanotubes (*8-10*). The earlier experiments using GO laminates in air with a typical $d ≈10$ Å have been explained by assuming one monolayer of moving water. For GO laminates soaked in water, $d$ increases to ≈13±1 Å, which allows two or three monolayers (*19,22,23,29*). Taking into account the effective thickness of graphene of 3.4 Å (interlayer distance in graphite), this yields a pore size of ≈9-10 Å, in agreement with the mesh size found experimentally.

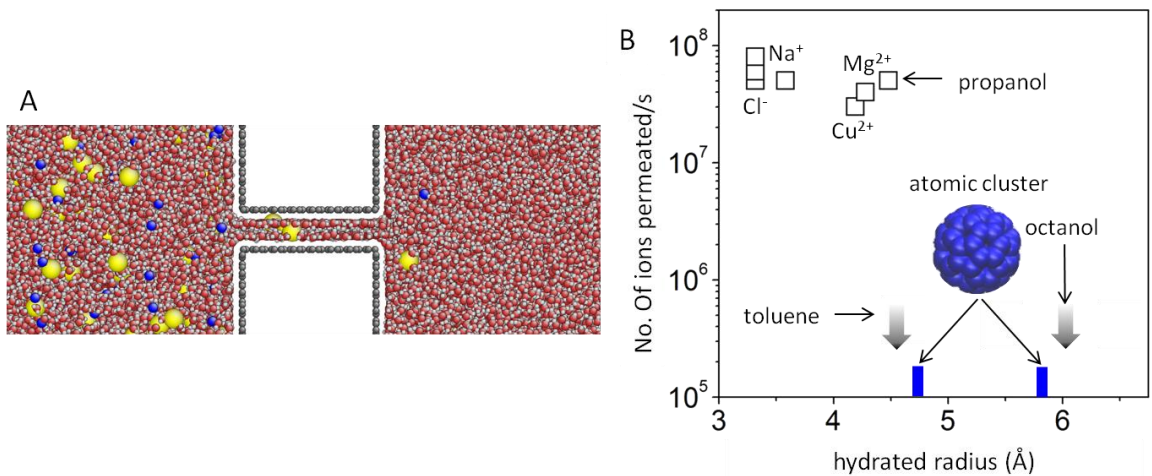

*Fig. 3. Simulations of molecular sieving. (**A**) Snapshot of NaCl diffusion through a 9 Å graphene slit allowing two monolayers of water. $Na^+$ and $Cl^-$ ions are in yellow and blue, respectively. (**B**) Permeation rates for NaCl, $CuCl_2$, $MgCl_2$, propanol, toluene and octanol for capillaries containing two monolayers of water. For octanol poorly dissolved in water, the hydrated radius is not known and we use its molecular radius. Blue marks: Permeation cutoff for an atomic cluster (pictured in the inset) for capillaries accommodating two and three monolayers of water (width of 9 Å and 13 Å, respectively).*

To support our model, we have used molecular dynamics simulations (MDS). The setup is shown in Fig. 3A where a graphene capillary separates feed and permeate reservoirs, and its width is varied between 7 and 13 Å to account for the possibility of one, two or three monolayers of water (*25*). We find that the narrowest MDS capillaries become filled with a monolayer of ice as described previously (*11*) and do not allow inside even such small ions as $Na^+$ and $Cl^-$. However, for two and three monolayers expected in the fully hydrated state, ions enter the capillaries and diffuse into the permeate reservoir. Their permeation rates are found approximately the same for all small ions and show little dependence on ionic charge (Fig. 3B). Larger species (toluene and octanol) cannot permeate even through capillaries containing three monolayers of water (*Fig. S3*). We have also modeled large solutes as atomic clusters of different size (*25*) and found that the capillaries accommodating 2 and 3 water monolayers rejects clusters with the radius larger than ≈4.7 and 5.8 Å, respectively. This probably indicates that the ion permeation through GO laminates is limited by regions containing two monolayers of water. The experimental and theory results in Figs 2 & 3B show good agreement.

Now we turn our attention to the absolute value of ion permeation rates found experimentally. Following ref. 11, one can estimate that, for our laminates with $h$ ≈5 μm and $L$ ≈1 μm, the effective length of graphene capillaries is $L \times h/d$ ≈5 mm and they occupy $d/L$ ≈0.1% of the surface area of the GO membrane. Accordingly, for a typical diffusion coefficient of ions in water (≈$10^{-5}$ $cm^2/s$), we expect permeation rates thousands times smaller than those found experimentally. Moreover, this estimate neglects the fact that functionalized regions narrow the effective water column (*11*). To appreciate how fast the observed permeation is, we have used the standard coffee filter paper and found the same diffusion rates for the paper of 1 mm in thickness (the diffusion barrier is equivalent to a couple of mm of pure water). Such fast transport of small ions cannot be explained by the confinement, which increases the diffusion coefficient by 50%, reflecting the change from bulk to two-dimensional water (*25*). Furthermore, functionalized regions [modeled as graphene with randomly attached epoxy groups (*20,21*)] do not enhance diffusion but rather suppress it (*25,29*) as expected because of the broken translational symmetry.

To understand the ultrafast ion permeation, we recall that graphene and GO powders exhibit a high adsorption efficiency with respect to many salts (*30*). Despite being very densely stacked, GO laminates are found to retain this property for salts with small hydrated radii [section 5 of (*25*)]. Our experiments show that permeating salts are adsorbed in amounts reaching as much as 25% of membranes' initial weight (*Fig. S2*). The large intake implies highly concentrated solutions inside graphene capillaries (close to the saturation). Our MDS simulations confirm that small ions prefer to reside inside capillaries (*Fig. S4*). The affinity of salts to graphene capillaries indicates an energy gain with respect to the bulk water, and this translates into a capillary-like pressure that acts on ions within a water medium, rather than on water molecules in the standard capillary physics (*25*). Therefore, in addition to the normal diffusion, there is a large capillary force, sucking small ions inside the membranes and facilitating their permeation. Our MDS provide an estimate for this ionic pressure as ≈50 bars (*25*).

The reported GO membranes exhibit extraordinary separation properties and their full understanding will require further work both experimental and theoretical. In particular, it remains unclear whether the functionalized regions play any other significant role, except for being spacers. In fact, the regions are negatively charged in water and may further enhance the ion sponge effect with respect to pristine graphene capillaries. With the ultrafast ion transport and atomic-scale pores, GO membranes present an interesting material to consider for separation and filtration technologies, particularly, those that target extraction of valuable solutes from complex mixtures.

## Supplementary Materials

### 1. Fabrication and characterization of GO membranes

Graphite oxide was prepared by exposing millimeter size flakes of natural graphite to concentrated sulfuric acid, sodium nitrate and potassium permanganate (Hummers' method) (*31*). Then, graphite oxide was exfoliated into monolayer flakes by sonication in water, which was followed by centrifugation at 10,000 rpm to remove remaining few-layer crystals. GO membranes were prepared by vacuum filtration of the resulting GO suspension through Anodisc alumina membranes with a pore size of 0.2 μm. By changing the volume of the filtered GO solution, we could accurately control the thickness *h* of the resulting membranes, making them from 1 to more than 10 μm thick. For consistency, all the membranes described in this report were chosen to be 5 μm in thickness, unless a dependence on *h* was specifically investigated.

We usually left GO laminates on top of the Anodiscs that served as a support to improve mechanical stability. In addition, we have checked influence of this porous support on permeation properties of GO and found them similar to those of free standing membranes (*11*). Finally, the GO membranes were glued onto a copper foil to cover an aperture of typically 1 cm in diameter (see Fig. 1 of the main text). The Cu foil was then clamped between O-rings, which provided a vacuum-tight seal and separated two containers that we refer to as feed and permeate.

Membranes were thoroughly tested for any possible cracks or holes by using a helium-leak detector as described in ref. 11. We believe this test is very important for the following reasons. During the last two years, four groups (*15-18*) studied filtration properties of GO laminates and, although results varied widely due to different fabrication and measurement procedures, they reported appealing characteristics including large water fluxes and notable rejection rates for certain salts. Unfortunately, large organic molecules were also found to pass through such GO filters (*16-18*). The latter observation is disappointing and would considerably limit interest in GO laminates as molecular sieves. However, the observation is inconsistent with the known structure of GO laminates (*20,21*). Small ions can reasonably be expected to move between GO planes separated by $d \leq 13\pm1$Å (*19,22*) but large molecules that do not fit within the available interlayer space must not. In this respect, we note that the emphasis of the previous studies was on high water rates that could be comparable or exceed the rates used for industrial desalination. Accordingly, a high water pressure was applied (*16-18*) and the GO membranes were intentionally prepared as thin as possible, 10–50 nm thick (*16,17*). This thickness corresponds to only a few dozens of GO layers that are made of randomly stacked μm-sized crystallites rather than being continuous graphene sheets. We speculate that such thin stacks contained holes and cracks (some may appear after applying pressure), through which large organic molecules could penetrate.

To check the laminar structure of our GO membranes, we performed X-ray diffraction measurements, which yielded the interlayer separation *d* of 9–10 Å at a relative humidity of 50±10%, in agreement with the previous studies (*11,22*).

### 2. Monitoring ion diffusion by electrical measurements

For a quick qualitative test of ion permeation through GO membranes, we have used the setup shown in the inset of Fig. S1. The feed container is initially filled with a concentrated solution of a tested salt, and the permeate container with deionized water. A typical feed solution is approximately a million times more electrically conducting than deionized water at room temperature. Therefore, if ions diffuse through the membrane, this should result in an increase in conductivity of water at the permeate side (*15*). Permeation of salts in concentrations at a sub-μM level can be detected in this manner.

Figure S1 shows examples of our measurements for the case of NaCl and potassium ferricyanide $K_3[Fe(CN)_6]$. The observed decreasing resistivity as a function of time indicates that NaCl permeates through the membrane. Similar behavior is observed for $CuSO_4$, KCl and other tested salts with small ions (see the main text). On the other hand, no noticeable changes in conductivity of deionized water can be detected for a potassium ferricyanide solution during measurements lasting for many days (Fig. S1).

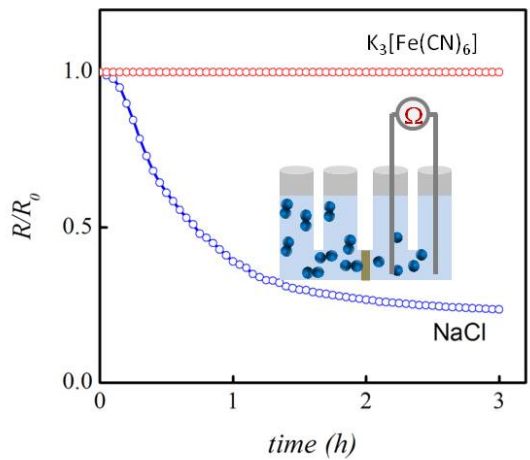

Fig. S1. Permeation of salts through GO membranes can be detected by using electrical measurements. The inset shows the measurement setup, and the main figure plots relative changes in resistivity of water with time in the permeate container. Changes are normalized to an initial value of measured resistance of deionized water.

### 3. Quantitative analysis of ion and molecular permeation

The above electrical measurements qualitatively show that small ions can permeate through our GO membranes whereas large ions such as $[Fe(CN)_6]^{3-}$ cannot. The technique is not applicable for molecular solutes because they exhibit little electrical conductivity. To gain quantitative information about the exact amount of permeating ions as well as to probe permeation of molecular solutes, we have carried out chemical analysis of water at the permeate side. Samples have been taken at regular intervals from a few hours to a few days and, in some cases, after several weeks, depending on a solute.

The ion chromatography (IC) and the inductively coupled plasma optical emission spectrometry (ICP-OES) are the standard techniques used to analyze the presence of chemical species in solutions (*32-34*). We have employed the IC for anionic species, and the ICP-OES for cations. In addition, we have crosschecked these measurements by weighing a dry material left after evaporation of water in the permeate container, which also allows us to find the amount of the salt permeated through GO membranes. Results of the weight and chemical analyses are in quantitative agreement.

The optical absorption spectroscopy has been employed in the case of larger ions such as $[Fe(CN)_6]^{3-}$, $[Ru(bipy)_3]^{2+}$ of Tris(bipyridine)ruthenium(II) dichloride ($[Ru(bipy)_3]Cl_2$) and $PTS^{4-}$ of pyrenetetrasulfonic acid tetrasodium salt ($Na_4PTS$). The latter technique is also widely used to detect solutes with absorption lines in the visible spectrum (*1,12*). We have observed no signature of $[Fe(CN)_6]^{3-}$, $[Ru(bipy)_3]^{2+}$ and $PTS^{4-}$ on the permeate side, even after many weeks of running the analysis.

To detect organic solutes such as glycerol, sucrose and propanol, we have employed the total organic carbon (TOC) analysis (*34,35*). No traces of glycerol and sucrose could be found in our permeate samples after several weeks, but propanol could permeate, although at a rate much lower than small ions as shown in Fig. 2 of the main text. The detection limit of our TOC is about 50 µg/L, and this puts an upper limit on permeation of the solutes that could not be detected in the permeate container. The corresponding limiting values are shown by arrows in Fig. 2 of the main text.

**4. Tested solutes and their hydrated radii**

Table S1. List of analyzed species and their hydrated radii.

| Hydrated radius (Å) | Ion/molecule |
|---|---|
| 3.31 | $K^+$ |
| 3.32 | $Cl^-$ |
| 3.58 | $Na^+$ |
| 3.75 | $CH_3COO^-$ |
| 3.79 | $SO_4^{2-}$ |
| 3.85 | $AsO_4^{3-}$ |
| 3.94 | $CO_3^{2-}$ |
| 4.19 | $Cu^{2+}$ |
| 4.28 | $Mg^{2+}$ |
| 4.48 | propanol |
| 4.65 | glycerol |
| 4.75 | $[Fe(CN)_6]^{3-}$ |
| 5.01 | sucrose |
| 5.04 | $(PTS)^{4-}$ |
| 5.90 | $[Ru(bipy)_3]^{2+}$ |

Figure 2 of the main text summarizes the results of our experiments by plotting the observed permeation rates as a function of the hydrated radius for all the solutes that have been analyzed quantitatively in our work. Values of the hydrated radius for eleven of them can be found in the literature (*13,26,36,37*). To the best of our knowledge, there exist no literature values for propanol, sucrose, glycerol and $PTS^{4-}$. In the latter case, we have estimated their hydrated radii by using their Stokes/crystal radii (*26,38*). To this end, we have plotted the known hydrated radii as a function of the known Stokes radii, which yields a simple linear dependence. Hydrated radii for the remaining 4 species can then be estimated by using this dependence and the corresponding Stokes radii. The resulting hydrated radii are listed in Table 1.

In principle, one can also consider using the Stokes rather than hydrated radius as a running parameter in Fig. 2 of the main text. In this case, the general trend of blocking large molecules and allowing small ones remains the same. However, glycerol and $[Fe(CN)_6]^{3-}$ that are blocked by GO membranes have Stokes radii smaller than those of permeating ions $Cu^{2+}$ and $Mg^{2+}$. Therefore, the functional form in Fig. 2 would be no longer single-valued near the cutoff, which is why we have chosen heuristically and for the physics reasons to use the hydrated radius.

**5. Ion sponging by GO membranes**

To test the adsorbing efficiency of GO laminates with respect to salts, we have carried out the following experiments. Our GO membranes were accurately weighed and placed in a solution with salt's concentration $C$ (we used $MgCl_2$, KCl and $K_3[Fe(CN)_6]$). After several hours, the laminates were taken out, rinsed with deionized water and dried. An intake of the salts was then measured. Fig. S2 shows that for salt that cannot permeate through GO laminates, there is no increase in weight. On the other hand, for the salts that fit inside GO capillaries, we have observed a massive intake that reaches up to 25% in weight (Fig. S2). The intake rapidly saturates at relatively small $C$ of ≈0.1 M. Further analysis shows that more than a half of this intake is a dry salt with the rest being additional bound water.

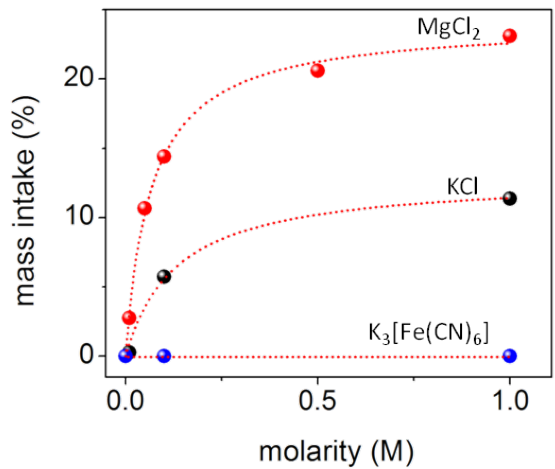

Fig. S2. Salt intake by GO membranes: Relative increase in weight for 5μm thick laminates soaked in different solutions. No intake could be detected for $K_3[Fe(CN)_6]$ but it is large for small-radius salts. All the weighing was carried out at the same relative humidity of 40±10 %.

The large salt intake proves that the permeating solutes accumulate inside GO capillaries, leading to highly concentrated internal solutions. Using the measured amount of adsorbed salts and the known amount of water in fully hydrated GO laminates, we estimate that concentrations of the internal solutions reach several molar, that is, can exceed external $C$ by a factor of 10 or more. This "salt sponge" effect is in qualitative agreement with the strong adsorption properties reported previously for graphene and GO powders (*30*).

The accumulation of salts inside GO capillaries means that there is a significant energy gain when ions move inside capillaries from the bulk solution (*39*). Our MDS below confirm this effect and indicate that the energy gain is mostly due to interaction of ions with graphene walls. The ion sponging is reminiscent of the standard capillary effects where molecules can gain energy by moving inside confined regions. In our case, water plays a role of a continuous medium in which the capillary-like pressure acts on ions, sucking them inside capillaries from the bulk water.

## 6. Molecular dynamics simulations

Our basic modeling setup consisted of two equal water reservoirs connected by a capillary formed by parallel graphene sheets as shown in Fig. 3A of the main text. Sizes of the reservoirs and capillaries varied in different modeling experiments. To analyze the salt-sponge effect and study ion diffusion in the confined geometry, we used reservoirs with a height of 51.2 Å, a length of 50 Å and a depth of 49.2 Å, which were connected by a 30 Å long capillary. A slightly smaller setup was used to assess sieving properties of graphene capillaries. It consisted of the reservoirs with a height of 23.6 Å, a length of 50 Å and a depth of 30.1 Å, which were connected by a 20 Å long capillary. For both setups, we varied the capillary width $d$ from 7 to 13 Å ($d$ is the distance between the centers of the graphene sheets). When the same property was modeled, both setups yielded similar behavior.

Periodic boundary conditions were applied in the Z direction, that is, along the capillary depth. Ions or molecules were added until the desired molar concentrations were reached. Water was modeled by using the simple point charge model (*40,41*). Sodium and chlorine ions were modeled by using the parameters from Refs 42-43; magnesium and copper anions with the OPLS-AA parameters (*44*). Intermolecular interactions were described by the 12-6 Lennard-Jones (LJ) potential together with a Coulomb potential. Parameters for water/graphene interactions were reported in Refs 45-46.

The system was initially equilibrated at 300 K with a coupling time of 0.1 $ps^{-1}$ for 500 ps (*47*). In the modeling of sieving properties, our typical simulation runs were 100 ns long and obtained in the isobaric ensemble at the atmospheric pressure where the simulation box was allowed to change only in the X and Y direction with a pressure coupling time of 1 $ps^{-1}$ (*45*) and a compressibility of $4.5 \times 10^{-5}$ $bar^{-1}$. The cutoff distance for nonbonding interactions was set up at 10 Å, and the particle mesh Ewald summations method was used to model the system's electrostatics (*48*). During simulations, all the graphene atoms were held in fixed positions whereas other bonds were treated as flexible. A time step of 1 fs was employed.

To model sieving properties of graphene, the GROMACS software was used (*49*). At the beginning of each simulation run, water molecules rapidly filled the graphene capillary forming one, two or three layer structures, depending on *d*, in agreement with the previous reports (*11,29*). Then after a certain period of time, which depended on a solute in the feed reservoir, ions/molecules started enter the capillary and eventually reached the pure water reservoir for all the modeled solutes, except for toluene and octanol. The found permeation rates are shown in Fig. 3B of the main text. We have also noticed that cations and anions move through the capillary together and without noticeably changing their hydration shells.

**7. Theoretical analysis of permeation for large molecules**

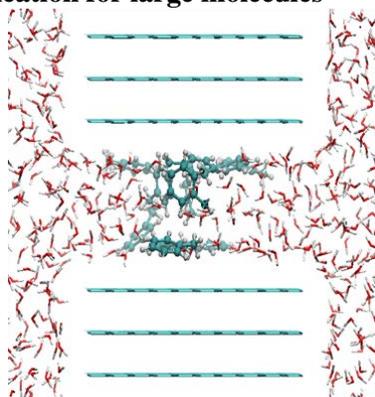

Fig. S3. Snapshot of our molecular dynamics simulations for toluene in water. All toluene molecules are trapped inside the short graphene channel and none leaves it even after 100 ns.

In the case of organic molecules (for example, propanol) our simulations showed that they entered the graphene capillary but then rapidly formed clusters that resided inside the capillary for a long time. The cluster formation is probably due to confinement. On the other hand, the long residence times can be attributed to van der Waals forces between the alcohol molecules and graphene (*50*). Toluene molecules exhibited even stronger interaction with graphene (due to π-π staking). In our simulations, toluene molecules entered the channel but never left it being adsorbed to graphene walls (Fig. S3). This adsorption is likely to be responsible for the experimentally undetectable level of toluene permeation. Therefore, despite our experimental data exhibit a rather simple sieving behavior that can be explained just by the physical size effect, we believe that van der Waals interactions between solutes and graphene may also play a role in limiting permeation for those molecules and ions that have sizes close to the cutoff radius.

To better understand the observed sieving effect with its sharp physical cutoff, we performed the following analysis. An artificial cluster was modeled as a truncated icosahedron and placed in the middle of the capillary as shown in the inset of Fig. 3B of the main text. The size of the cluster was varied by changing the distance between the constituent 60 atoms, and the interaction energy between the cluster and the graphene capillary was calculated. The energy was computed as the sum of interactions between all the atoms involved which were modeled with a 12-6 LJ potential. Positive and negative values of the calculated energy indicate whether the presence of the cluster in the capillary is energetically favorable or not, respectively. The minimum radius for which the spherical cluster was allowed into the graphene capillary obviously depended on the capillary size. For capillaries that allowed two monolayers of waters ($d = 9$Å) this radius was found to be 4.7Å. For wider capillaries containing three water monolayers ($d = 13$Å), the radius was 5.8Å. These values are shown in Fig. 3B of the main text as the blue bars.

## 8. Simulations of the ion sponge effect

In this case, we employed a relatively long capillary (482 Å) such that its volume was comparable to that of the reservoirs (see Fig. S4A). The capillary width was varied between 9 and 11 Å, which corresponds to 2 and 3 monolayers of water. MDS were carried out in a canonical ensemble using LAMMPS (*51*). The temperature was set at 300 K by using the Nose-Hoover thermostat. The equations of motion were integrated using a velocity-Verlet scheme with a time step of 1.0 fs. The snapshots obtained in these simulations (an example is shown in Fig. S4A) were processed by Atomeye (*52*).

During the simulations, we counted the number of ions inside the capillary as a function of time (Fig. S4B). If the initial concentration $C$ of NaCl was taken constant over the entire system (for example, $C$ =1 M for the black curve in Fig. S4B), we found that the salt moved from the reservoirs into the capillary, that is, ions were attracted to the confined region. Then, we used smaller initial $C$ inside the two reservoirs (0.1 and 0.5 M) while keeping the same $C$ =1 M inside the capillary. Despite the large concentration gradient, the salt still moved into the capillary rather than exited it (see Fig. S4B).

In the next MDS experiment, we kept a low concentration of NaCl in the two reservoirs ($C$ =0.1 M) and gradually increased $C$ inside the graphene capillary up to 3 M. For $C$ =2 M inside it, we still observed an influx of NaCl from the reservoirs. The salt started leaving the capillary only if $C$ inside approached ~3 M. This allows an estimate of the equilibrium concentration of NaCl inside the graphene capillary as 2–3M, in good agreement with the experiments discussed in section #5. The concentration gradient corresponds to a capillary-like pressure of ≈50 bars, which acts on salt ions against the osmotic pressure. To find an average speed of ion flow caused by this ionic pressure, further work is required, especially because the viscosity experienced by ion moving against graphene walls is likely to be different from that for water molecules.

We have also assessed whether functionalized GO regions can play any major role in the salt sponge effect and, more generally, in molecular permeation through GO laminates. To this end, we used the same MDS setups as described above but added hydroxyl and epoxy groups to both walls of graphene capillaries. The epoxy group was modeled by binding an oxygen atom to two carbon atoms of graphene and the hydroxyl group (OH) by its oxygen bonded to a carbon atom. For simplicity, oxygen atoms were fixed in their positions whereas the O-H bond was treated as flexible. Fig. S4C shows an example of the latter simulations. Both ion and water dynamics inside GO capillaries is found to be extremely slow, in agreement with the assumptions of our previous work (*11*) and a recent MDS report (*29*). Accordingly, we expect that the sponge effect should be weaker for functionalized capillary regions compared to pristine ones.

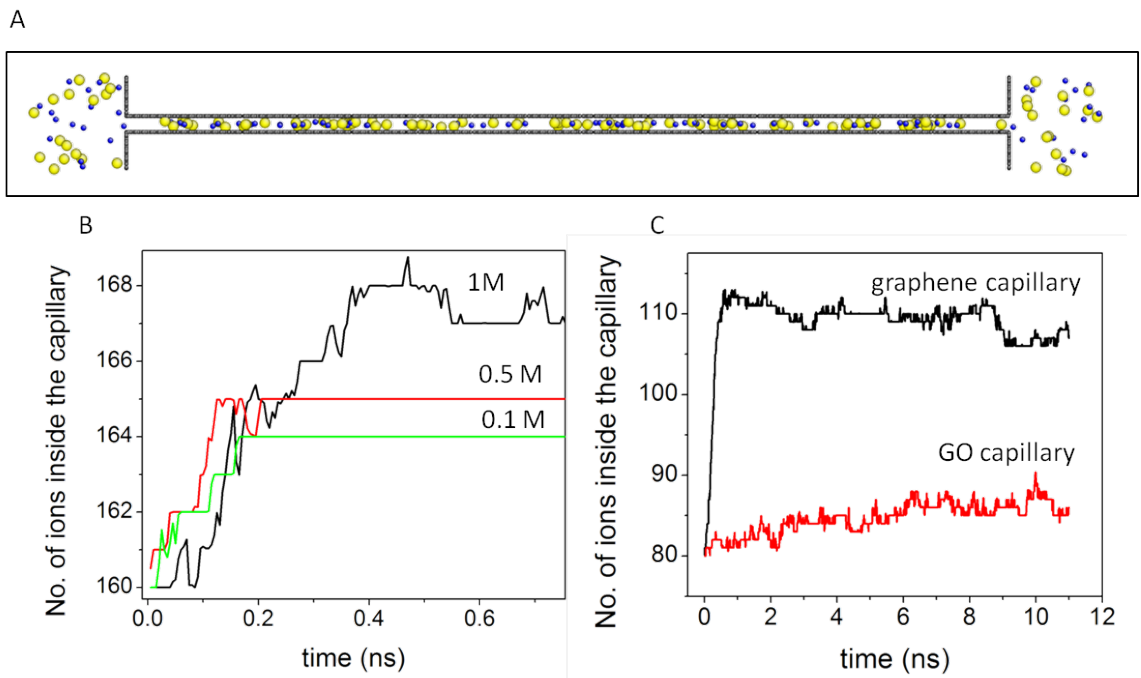

Fig. S4. Simulated salt-sponge effect. (**A**) Snapshot for the case of 1M NaCl inside the capillary and 0.1 M in the reservoirs (water molecules are removed for clarity). Despite the concentration gradient, ions move from the reservoirs into the capillary. (**B**) Number of ions inside a 9 Å wide capillary (two monolayers of water) as a function of simulation time. Initial concentrations of NaCl in the two reservoirs were 0.1, 0.5 and 1 M for the different curves. The initial NaCl concentration inside the capillary was the same for all the curves ($C$ =1 M). (**C**) Comparison between graphene and GO capillaries. Evolution of the number of ions inside a capillary (11Å wide) for initial $C$ =1 M throughout the system.